\documentclass{appolb_edited}

\usepackage{amsmath}
\usepackage{amsfonts}
\usepackage{amssymb}
\usepackage{mathrsfs}
\usepackage{hyperref}
\usepackage{xspace}
\usepackage{bookmark}

\usepackage[
	bibstyle=phys,
	biblabel=brackets,
	citestyle=numeric-comp,
	sorting=none,
	doi=false,
	eprint=true,
	pageranges=false,
	backend=biber
]{biblatex}

\DeclareFieldFormat[article]{title}{\mkbibitalic{#1\isdot}}
\AtEveryBibitem{
	\ifentrytype{article}{
		\clearfield{title}
		\clearfield{eprint}
	}{}
	\ifentrytype{online}{
		\clearfield{title}
		\clearfield{year}
	}{}
	\ifentrytype{book}{
		\clearfield{series}
		\clearfield{volume}
	}{}
}

\newcommand{\ket}[1]{|#1\rangle}

\newcommand{\updown}[2]{^{#1}_{\phantom{#1}#2}}
\newcommand{\downup}[2]{_{#1}^{\phantom{#1}#2}}

\newcommand{\D}{{\cal D}}
\newcommand{\R}{{\mathscr R}}
\newcommand{\Rt}{\,{}^{(3)}\!R}
\newcommand{\Et}{\widetilde E}
\newcommand{\de}{{\rm det}\,E}
\newcommand{\Gc}{{\Gamma_{\! 0}}}

\begin{document}

\title{SCALAR CURVATURE OPERATOR FOR LOOP QUANTUM GRAVITY ON A CUBICAL GRAPH\thanks{Presented at the 8$^{\rm th}$ Conference of the Polish Society on Relativity, Warsaw, Poland, 19--23 September 2022.}%
}
\author{Ilkka M\"akinen
	\address{National Centre for Nuclear Research, Pasteura 7, 02-093 Warsaw, Poland \\[1ex]
	Faculty of Physics, University of Warsaw, Pasteura 5, 02-093 Warsaw, Poland \\[1ex]
	\texttt{ilkka.makinen@ncbj.gov.pl}}
}

\maketitle

\begin{abstract}
	We introduce a new operator representing the three-dimensional scalar curvature in loop quantum gravity. The operator is constructed by writing the Ricci scalar classically as a function of the Ashtekar variables and regularizing the resulting expression on a cubical spin network graph. While our construction does not apply to the entire Hilbert space of loop quantum gravity, the proposed operator can be applied to concrete calculations in various approaches which are derived from the framework of full loop quantum gravity using states defined on cubical graphs.
\end{abstract}

\section{Introduction}

Loop quantum gravity (see \eg \cite{Ashtekar:2017yom, Rovelli:2004tv, Thiemann:2007pyv}) is one of the main approaches to the problem of quantum gravity, providing a concrete realization of a quantum theory of gravity as a theory of quantum geometry. Accordingly, a key role in the theory is played by quantum operators representing geometrical quantities such as volumes, areas, lengths and angles. Another example is the three-dimensional Ricci scalar (in the setting of a 3+1 decomposition of general relativity), which is relevant to loop quantum gravity both as a fundamental geometrical observable characterizing the curvature of the spatial manifold, as well as a possible ingredient for the dynamics of the theory. Indeed, the Hamiltonian constraint of general relativity can be expressed in the Ashtekar variables \cite{Ashtekar:1987gu, BarberoG:1994eia} as
\begin{equation}
	C = \frac{1}{\beta^2}\frac{\epsilon\updown{ij}{k}E^a_iE^b_jF_{ab}^k}{\sqrt{|\de|}} + (1+\beta^2)\sqrt{|\de|}\Rt,
	\label{C_new}
\end{equation}
with the scalar curvature replacing the expression of the Lorentzian term in terms of the extrinsic curvature, whose use in loop quantum gravity was popularized by the pioneering work of Thiemann \cite{Thiemann:1996aw}.

An operator representing the scalar curvature in loop quantum gravity has been introduced previously in \cite{Alesci:2014aza}. The construction is based on the ideas of Regge calculus, where the smooth physical manifold is approximated by a fictitious auxiliary manifold of singular geometry, where curvature is concentrated entirely on one-dimensional line segments. In our article \cite{Lewandowski:2021iun} we propose a much more direct approach towards the quantization of the Ricci scalar. To bypass a certain technical difficulty, which will be touched upon in Section \ref{regularization}, we do not attempt to define our operator on the entire Hilbert space of loop quantum gravity. Instead, the operator is constructed on the space of states defined on a fixed cubical spin network graph. As such, our operator can be applied to calculations in models such as quantum-reduced loop gravity \cite{Alesci:2013xd, Alesci:2016gub, Makinen:2020rda} and effective dynamics \cite{Dapor:2017rwv, Han:2019vpw, Zhang:2021qul}, which are derived from the formalism of loop quantum gravity using states defined on cubical graphs. Moreover, the framework of algebraic quantum gravity \cite{Giesel:2006uj, Giesel:2007wn} has shown how a mathematically complete quantization of the gravitational field can be achieved entirely in terms of states defined on a single cubical graph.

\section{Classical preparations}

The classical object which we wish to promote into an operator in loop quantum gravity is the three-dimensional Ricci scalar integrated over the spatial manifold, \ie
\begin{equation}
	\int d^3x\,\sqrt q\Rt.
	\label{int qR}
\end{equation}
Our construction begins by expressing the integrand in \eqref{int qR} directly in terms of the Ashtekar variables. In the metric formulation, the Ricci scalar is given by the expression
\begin{equation}
	\Rt = q^{ab}\bigl(\partial_c\Gamma^c_{ab} - \partial_b\Gamma^c_{ac} + \Gamma^c_{ab}\Gamma^d_{cd} - \Gamma^c_{ad}\Gamma^d_{bc}\bigr)
	\label{R(q)}
\end{equation}
where $\Gamma^a_{bc}$ are the Christoffel symbols corresponding to the spatial metric $q_{ab}$. The spatial metric is related to the densitized triad $E^a_i$ by
\begin{equation}
	q^{ab} = \frac{E^a_iE^b_i}{|\de|}.
	\label{q^ab}
\end{equation}
Inserting \Eq{q^ab} into \Eq{R(q)} and carrying out a straightforward (if rather lengthy) calculation, we obtain an expression for the Ricci scalar as a function of the densitized triad and its first and second derivatives. For our present discussion, we express this relation symbolically as
\begin{equation}
	\sqrt q\Rt = \R\bigl(E^a_i, \, \partial_aE^b_i, \, \partial_a\partial_bE^c_i\bigr).
	\label{R(E)}
\end{equation}
The explicit expression of the function $\R$ is reported in \cite{Lewandowski:2021iun}. 

The function $\R\bigl(E^a_i, \, \partial_aE^b_i, \, \partial_a\partial_bE^c_i\bigr)$ is not manifestly invariant under the $SU(2)$ gauge transformations corresponding to internal rotations of the densitized triad, due to the non-covariant transformation properties of the partial derivatives of the triad. Consequently, it would be difficult to ensure that a gauge invariant curvature operator is obtained if the Ricci scalar is quantized on the basis of \Eq{R(E)}. A more appropriate starting point for quantization can be found by replacing the partial derivatives of the triad with the gauge covariant derivatives defined by 
\begin{equation}
	\D_aE^b_i = \partial_aE^b_i + \epsilon\downup{ij}{k}A_a^jE^b_k
	\label{DE}
\end{equation}
where $A_a^i$ is the Ashtekar connection. Under a local $SU(2)$ gauge transformation described by a gauge function $g(x)\in SU(2)$, the matrix-valued variable $\D_aE^b = \D_aE^b_i\tau^i$ transforms as $\D_aE^b(x) \to g(x)\D_aE^b(x)g^{-1}(x)$.

When \Eq{DE} is used to express the partial derivatives in \Eq{R(E)} in terms of the gauge covariant derivatives, a direct calculation shows that the partial derivatives can be substituted with covariant derivatives ``for free'' (the terms proportional to the connection $A_a^i$ cancel out among themselves) provided that the second partial derivative $\partial_a\partial_bE^c_i$ is replaced with the symmetric part $\D_{(a}\D_{b)}E^c_i$ of the second covariant derivative. That is, the Ricci scalar is the {\em same} function of the triad and its gauge covariant derivatives, as of the triad and its partial derivatives:
\begin{equation}
	\sqrt q\Rt = \R\bigl(E^a_i, \, \D_aE^b_i, \, \D_{(a}\D_{b)}E^c_i\bigr).
	\label{R(DE)}
\end{equation}
Thanks to the covariant transformation law of the covariant derivatives, the right-hand side of \Eq{R(DE)} is manifestly $SU(2)$ gauge invariant, and it is this expression that we take as the classical starting point for the construction of the curvature operator.

\section{Regularization on a cubical graph}
\label{regularization}

Now the task is to turn the expression \eqref{R(DE)}, integrated over the spatial manifold, into an operator on the Hilbert space of loop quantum gravity. To accomplish this, the integral must be regularized by expressing it in terms of elements which correspond to well-defined operators in loop quantum gravity. The classical variables corresponding to the elementary operators of loop quantum gravity are holonomies (parallel propagators) of the Ashtekar connection along one-dimensional curves and fluxes of the densitized triad through two-dimensional surfaces, but as we will soon see, certain combinations of these operators can also turn out to be very useful.

The kinematical Hilbert space of loop quantum gravity is spanned by the so-called spin network states. A spin network state is labeled by a graph $\Gamma$ together with a spin quantum number $j_e$ for each edge of the graph and a $SU(2)$ tensor $\iota_v$ (of appropriate index structure) for each vertex\footnote{
	A distinction is often made between generalized spin network states, which carry arbitrary tensors at their vertices and span the entire kinematical Hilbert space, and proper spin network states, which are labeled by invariant tensors and span the gauge invariant Hilbert space (with respect to $SU(2)$ gauge transformations generated by the Gauss constraint operator).
}.
However, constructing a consistent regularization of the covariant derivatives of the triad on graphs of arbitrary, irregular shape is a complicated technical challenge, to which we have no satisfactory solution at the moment. We therefore restrict ourselves to the much more modest problem of defining the curvature operator on the Hilbert space of states based on a fixed cubical graph, \ie a graph whose vertices are six-valent, and whose edges are aligned with the coordinate directions of a fixed Cartesian background coordinate system.

To regularize the integrated scalar curvature on the lattice provided by the cubical graph, we partition the spatial manifold into cubical cells $\Box$, such that every cell contains a single vertex of the graph. For simplicity, we assume that each cell is a cube of coordinate volume $\epsilon^3$. For every vertex $v$ we introduce a family of three surfaces, denoted by $S^a(v)$ $(a = x, y, z)$, within the corresponding cell $\Box$. Each surface contains the vertex $v$ and is dual to the corresponding coordinate direction, \ie the coordinate $x^a$ is constant on the surface $S^a(v)$. The integrated Ricci scalar can then be approximated as a Riemann sum associated with the cubical partition:
\begin{equation}
	\int d^3x\,\sqrt q\Rt \simeq \sum_{\Box} \epsilon^3 \sqrt{q(v_\Box)}\Rt(v_\Box),
	\label{sum R}
\end{equation}
where $v_\Box$ denotes the vertex contained in the cell $\Box$.

When \Eq{R(DE)} is used to express the integrand in \Eq{sum R}, each instance of the densitized triad can be approximated by the flux variable
\begin{equation}
	E_i\bigl(S^a(v)\bigr) = \int_{S^a(v)} d^2\sigma\,n_aE^a_i,
	\label{}
\end{equation}
which satisfies $E_i\bigl(S^a(v)\bigr) = \epsilon^2 E^a_i(v) + {\cal O}(\epsilon^3)$ for small values of the regularization parameter $\epsilon$. As for the regularization of the covariant derivatives $\D_aE^b_i$, the appropriate technical tool is provided by the so-called parallel transported flux variable (also known as the gauge covariant flux in the literature). The parallel transported flux variable is defined by
\begin{equation}
	\Et(S, x_0) = \int_S d^2\sigma\,n_a(\sigma)h_{x_0, x(\sigma)}E^a_i\bigl(x(\sigma)\bigr)\tau^i h^{-1}_{x_0, x(\sigma)}
	\label{Et}
\end{equation}
where $h_{x_0, x(\sigma)}$ are holonomies of the Ashtekar connection, which connect each point $x(\sigma)$ on the surface $S$ to a fixed point $x_0$ (on the surface or outside of it) along a chosen system of paths $p_{x_0, x(\sigma)}$. Under a local $SU(2)$ gauge transformation, the parallel transported flux variable transforms covariantly at the point $x_0$: $\Et(S, x_0) \to g(x_0)\Et(S, x_0)g^{-1}(x_0)$.

For a given vertex $v$, let $v_a^-$ and $v_a^+$ denote the vertices which come before and after $v$ in the direction of the $x^a$-coordinate axis. Using the parallel transported flux variable, we construct the object
\begin{equation}
	\Delta_aE\bigl(S^b, v\bigr) = \frac{\Et\bigl(S^b(v_a^+), v\bigr) - \Et\bigl(S^b(v_a^-), v\bigr)}{2},
	\label{DaE}
\end{equation}
where the parallel transports to the central vertex $v$ are taken along the edges connecting $v_a^+$ and $v_a^-$ to $v$. The variable \eqref{DaE} represents a discrete approximation of the covariant derivative $\D_aE^b$ at $v$, corresponding to a symmetric discretization\footnote{
	The symmetric discretization is chosen in order to avoid introducing a preferred direction, since if one is given only a state defined on a cubical graph, there is no way to unambiguously determine which direction should be identified as the positive direction of any given coordinate axis.
}
of the form
\begin{equation}
	f'(x) \simeq \frac{f(x+\epsilon) - f(x-\epsilon)}{2\epsilon}.
	\label{f' sym}
\end{equation}
Expanding the right-hand side of \Eq{DaE} in powers of $\epsilon$, one finds
\begin{equation}
	\Delta_aE\bigl(S^b, v\bigr) = \epsilon^3\D_aE^b(v) + {\cal O}\bigl(\epsilon^4\bigr),
	\label{}
\end{equation}
confirming that the variable $\Delta_aE\bigl(S^b, v\bigr)$ indeed correctly approximates the covariant derivative of the triad. The same technique can be applied to regularize the second covariant derivatives, using the template
\begin{equation}
	f''(x) \simeq \frac{f(x+\epsilon) - 2f(x) + f(x-\epsilon)}{\epsilon^2}
	\label{f''}
\end{equation}
for the diagonal second derivatives $\D_a^2E^b$, and the symmetric discretization
\begin{align}
	\frac{\partial^2\!f(x,y)}{\partial x\partial y} \simeq \frac{1}{4\epsilon^2}\Bigl(&f(x+\epsilon, y+\epsilon) - f(x+\epsilon, y-\epsilon) \notag \\
	&- f(x-\epsilon, y+\epsilon) + f(x-\epsilon, y-\epsilon)\Bigr)
	\label{f_xy}
\end{align}
for the mixed second derivatives $\D_a\D_bE^c$. With a judicious choice of the paths involved in the parallel transported flux variables, one can arrange that the resulting discretized variable $\Delta_{ab}E\bigl(S^c, v\bigr)$ is symmetric in $a$ and $b$, and hence provides an approximation of the symmetric part of the second covariant derivative:
\begin{equation}
	\Delta_{ab}E\bigl(S^c, v\bigr) = \epsilon^4\D_{(a}\D_{b)}E^c(v) + {\cal O}\bigl(\epsilon^5\bigr).
	\label{}
\end{equation}

\section{The curvature operator}

The regularization of the integrated Ricci scalar is now completed by replacing the continuous variables in \Eq{R(DE)} with their discretized counterparts. This results in the regularized expression
\begin{equation}
	\int d^3x\,\sqrt q\Rt \simeq \sum_{\Box} \R\Bigl(E_i\bigl(S^a(v_\Box)\bigr), \, \Delta_aE_i\bigl(S^b, v_\Box\bigr), \, \Delta_{ab}E_i\bigl(S^c, v_\Box\bigr)\Bigr),
	\label{regularized}
\end{equation}
the discrete sum approximating the continuous integral in the limit of small regularization parameter. The factors of $\epsilon$ are precisely absorbed in the discretized variables with no factors left over, reflecting the fact that the integrand is geometrically a density of weight 1.

All of the discretized variables on the right-hand side of \Eq{regularized} correspond to a well-defined operators in loop quantum gravity. The integrated Ricci scalar can therefore be quantized simply by ``putting hats'' over these variables\footnote{
	Precisely speaking, we have neglected to discuss the quantization of the factors of $\det E$ appearing in the classical expression \eqref{R(DE)}. The treatment of these factors is standard, relying on techniques which are routinely used in the literature of loop quantum gravity. In particular, to account for the zero eigenvalues present in the spectrum of the volume operator, negative powers of the volume element $\sqrt{|\de|}$ are quantized using the Tikhonov-type regularized inverse
\[
	\widehat{{\cal V}_v^{-1}} \equiv \lim_{\delta\to 0} \widehat V_v\bigl(\widehat V_v^2 + \delta^2\bigr)^{-1}
\]
of the local volume operator $\widehat V_v$ (see \eg \cite{Bianchi:2008es}).
}.
When applied on a state in the Hilbert space of the fixed cubical graph $\Gamma_0$, the resulting operator takes the form
\begin{equation}
	\biggl(\widehat{\displaystyle\int d^3x\,\sqrt q\Rt}\biggr)\ket{\Psi_\Gc} = \sum_{v\in{\Gc}} \widehat{\R}_v\,\ket{\Psi_\Gc}
	\label{}
\end{equation}
where $\widehat\R_v$ denotes any symmetric factor ordering of the operator
\begin{equation}
	\R\Bigl(\widehat E_i\bigl(S^a(v)\bigr), \, \widehat{\Delta_aE_i}\bigl(S^b, v\bigr), \, \widehat{\Delta_{ab}E_i}\bigl(S^c, v\bigr)\Bigr).
	\label{}
\end{equation}

The operator $\widehat\R_v$ was studied in \cite{Lewandowski:2022xox} in the simplified kinematical setting of quantum-reduced loop gravity. In particular, we looked at expectation values of curvature in the standard basis states on the Hilbert space of the quantum-reduced model. We found that the expectation values tend to be markedly negative in a large class of states where one would {\em a priori} not expect either sign of the curvature to be strongly favoured. Since the states considered in our calculations lack any definite semiclassical interpretation, the physical significance of this result is not completely clear. However, on a technical level the negative expectation values can be traced back to the regularization of second derivatives represented by \Eq{f''}. If further calculations confirm that the problem of negative expectation values is encountered also in physically more realistic states, we expect that the problem could be resolved by using a modified discretization of second derivatives, where the central vertex $v$ is avoided altogether but one has to use four vertices instead of three to discretize the diagonal components of the second derivative.

\section{Conclusions}

We have proposed a new operator representing the three-dimensional scalar curvature in loop quantum gravity. The classical starting point of our work is to express the Ricci scalar directly as a function of the densitized triad and its gauge covariant derivatives. Due to difficulties associated with regularizing the covariant derivatives on arbitrary spin network graphs, we define our operator on the Hilbert space of a fixed cubical graph. From the perspective of full loop quantum gravity, the assumption of a cubical graph represents a significant limitation, and extending the construction to more general graphs is certainly an interesting question for future work. However, our construction is general enough to cover several physically motivated models of loop quantum gravity (quantum-reduced loop gravity, the effective dynamics approach) as well as algebraic quantum gravity, which provides a reformulation of loop quantum gravity in terms of states defined on a single cubical graph.

In the continuation article \cite{Lewandowski:2022xox} calculations were performed to probe the properties of the new curvature operator on the Hilbert space of quantum-reduced loop gravity. Our results indicate that the expectation values of curvature are consistently negative in certain states where one would intuitively think that neither sign of curvature should be clearly preferred over the other. This issue should be further clarified through a detailed semiclassical analysis of the curvature operator, in which one would study the peakedness properties of the operator with respect to semiclassical states peaked on given classical configurations (\eg a flat spatial geometry). If such calculations confirm that the expectation values of our operator are distributed too strongly towards the negative side, we expect that the problem could be resolved through a simple modification of the regularization of second covariant derivatives of the triad, as outlined in \cite{Lewandowski:2022xox}. 

\section*{Acknowledgments}

This work was funded by National Science Centre, Poland through grants no. 2018/30/Q/ST2/00811 and 2022/44/C/ST2/00023. For the purpose of open access, the author has applied a CC BY 4.0 public copyright license to any author accepted manuscript (AAM) version arising from this submission.

\section*{References}

\printbibliography[heading=none]

\end{document}